\documentclass[12pt]{iopart}
\usepackage{bm}
\usepackage{colordvi}
\usepackage{color}
\usepackage{graphicx}
\usepackage{textcase, url, ulem}

\begin{document}

\title[AB interferometry with a tunnel-coupled wire]{Aharonov-Bohm interferometry with a tunnel-coupled wire}

\author{A. Aharony$^{1, 2, 3}$, S. Takada$^4$, O. Entin-Wohlman$^{1, 2, 3}$, M. Yamamoto$^{4, 5}$ and S. Tarucha$^{4,6}$}

\address{$^1 $Department of Physics, Ben Gurion University, Beer
Sheva 84105, Israel}
\address{ $^2 $Ilse Katz Center for
Meso- and Nano-Scale Science and Technology, Ben Gurion
University, Beer Sheva 84105, Israel}
\address{$^3$ Raymond and Beverly Sackler School of Physics and Astronomy, Tel Aviv University, Tel Aviv 69978, Israel}
\address{$^4$ Department of Applied Physics, University of Tokyo, Bunkyo-ku, Tokyo, 113-8656, Japan}
\address{$^5$ PRESTO-JST, Kawaguchi-shi, Saitama, 332-0012, Japan}
\address{$^6$ Center for Emergent Matter Science (CEMS), RIKEN, Wako, Saitama, 351-0198, Japan}

\ead{aaharony@bgu.ac.il}

\begin{abstract}
Recent experiments [M. Yamamoto {\it et al.}, Nature Nanotechnology {\bf 7}, 247 (2012)] used the transport of electrons through an Aharonov-Bohm interferometer and two coupled channels (at both ends of the interferometer) to demonstrate a manipulable flying qubit. Results included in-phase and anti-phase Aharonov-Bohm (AB) oscillations of the two outgoing currents as a function of the magnetic flux, for strong and weak inter-channel coupling, respectively. Here we present new experimental results for a three terminal interferometer, with a tunnel coupling between the two outgoing wires. We show that in some limits, this system is an  even simpler realization of the `two-slit' experiment. We also present a simple tight-binding theoretical model which imitates the experimental setup. For weak inter-channel coupling, the AB oscillations in the current which is reflected from the device are very small, and therefore the oscillations in the two outgoing currents must cancel each other, yielding the anti-phase behavior, independent of the length of the coupling regime. For strong inter-channel coupling, whose range depends on the asymmetry between the channels, and for a relatively long coupling distance, all except two of the waves in the coupled channels  become evanescent. For the remaining running waves one has a very weak dependence of the  ratio between the currents in the two channels on the magnetic flux, implying that these currents are in phase with each other.
\end{abstract}

\maketitle

\section{Introduction}
In a recent paper, \cite{naturenano} some of us demonstrated a scalable flying qubit architecture in a four-terminal setup. Electrons were transported via an Aharonov-Bohm (AB) ring into two channel wires that have a tunable tunnel coupling between them.
 The superposition of the two electron states between the two outgoing channels can be considered as a flying qubit, which can be manipulated by the various gate voltages on the system.  That paper also exhibited an interesting variation of the relative phases of the AB oscillations in the currents in the two outgoing channels, as function of the coupling between them. In the present paper we present similar experimental  results for a new (simpler) three-terminal setup, and then present a simple theoretical model which reproduces them.

The early experiments on two-terminal `closed' AB interferometers \cite{phase} exhibited a phase rigidity: The minima and maxima of the AB oscillations in the outgoing current stayed at the same values of the magnetic flux through the interferometer ring, irrespective of the details of a quantum dot which was placed on one arm of the interferometer. At most, the phase of the oscillation jumped by $\pi$, interchanging the minima and maxima. This rigidity was due to the Onsager relation, by which unitarity and time reversal symmetry imply that the conductance through the interferometer must be an even function of the magnetic field. \cite{onsager}
One way to overcome this rigidity, and to measure the phase of the transmission amplitude through the quantum dot, was to open the interferometer, allowing leaks of electrons out of the interferometer ring and thus breaking the unitarity condition needed for the Onsager relation. \cite{us}
Indeed, experiments with open interferometers \cite{shuster} yielded a continuous shift in the oscillation phases. In a `two-slit' geometry (as in Young's classical diffraction experiment),  the electronic wave passes only once through each branch of the interferometer, and then this phase shift should be equal to the desired transmission phase. However, this `two-slit' limit is achieved only when the electronic leaks are very large, and therefore the remaining visibility (i.e. the amplitude of the AB oscillations) is very small. \cite{bih}

In the present paper we discuss an alternative way to cross between the `two-slit' and the `two-terminal' limits, i.e. between the case in which the oscillation phase reflects the scattering phase through one branch of the interferometer and the case of full phase rigidity.  This is achieved by having two outgoing wires, namely by our novel three-terminal setup, consisting of an AB ring and a coupled-wire. \cite{Tsukada}
A priori, the current in each outgoing wire need not obey the Onsager phase rigidity, because electrons ``leak" through the other outgoing wire. However, we show that the strength of the tunnel-coupling between the outgoing wires can cause the crossover between the above two limits.

 Section \ref{exp} describes our experimental setup, as shown in Fig. 1. This is similar to that of Ref. \cite{naturenano}, but now we use only three (and not four) terminals, one for the incoming current (on the left) and two for the outgoing coupled channels (on the right).
  Section \ref{exp} also presents results for the two outgoing currents, see Fig. 2. These results are also similar to those found in Ref. \cite{naturenano}. For strong inter-wire tunnel-coupling, the AB oscillations in both wires have the same phases, and they exhibit phase rigidity as in the `two-terminal' case. For weak tunnel coupling, the two oscillations have opposite phases (which we call `anti-phase'). In this limit the phases of the oscillations vary smoothly with the gate voltage on the upper branch of the interferometer, implying a relation between the measured phase and the phase of the electronic wave function on that branch.

\begin{figure}[hbtp]
\includegraphics{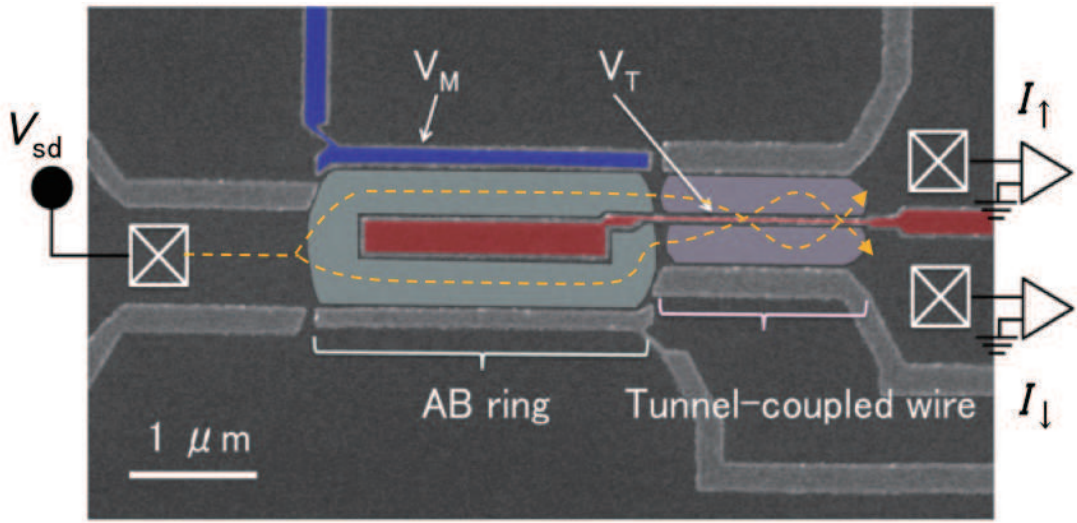}
\caption{The experimental setup for the interferometer (AB ring) plus the tunnel-coupled wire. The regions beneath the gate electrodes are depleted, so that the interferometer paths are defined in between. The tunnel coupling between the upper and lower wires is modulated by the narrow tunneling gate $V^{}_{\rm T}$. We apply the source-drain bias $V^{}_{\rm sd}$ on the left ohmic contact and measure the output currents of the two right ohmic contacts.}
\label{y2}
\end{figure}

In the rest of this paper we 
develop a simple and minimal theoretical tight-binding model, which imitates the experimental setup.
Since the following sections are somewhat technical, we first give a qualitative description of our model and of our results.
In Sec.  \ref{models} we formulate the tight binding equations which describe the scattering of electrons from two quantum network models, which are constructed from one-dimensional wires. In these models, which are shown in Fig. \ref{sAB}, the AB loop is modeled by a triangle (ABC) of one-dimensional wires, and the tunnel-coupling between the outgoing wires is modeled by many transverse wires, each having a tunneling energy $V$. These equations are solved for Fig. \ref{sAB}(a) in Sec. \ref{3term}, and for Fig. \ref{sAB}(b) in Secs. \ref{wires} and \ref{Intwires}. The amplitudes and the relative phases of the AB oscillations in the  reflected and transmitted currents depend on $V$. At small $V$, the oscillations in the reflected current are very small, and therefore
the oscillations in the two outgoing currents must cancel each other, hence the ``anti-phase" behavior. This phenomenon appears for both models in Fig. \ref{sAB}, independent of the length of the coupling between the outgoing wires. In contrast, the ``in-phase" behavior appears only for a long region of tunnel-coupling between the wires, Fig. \ref{sAB}(b). For a very large $V$, all the wave functions on these wires become evanescent, due to a strong reflection from each of the transverse coupling wires, and therefore both outgoing currents become very small. However, for intermediate values of $V$, in a range which depends on some anisotropy between the two wires, one finds only one ``running" wave solution on the coupled wires. For this single solution, the ratio of the wave amplitudes, and therefore also the ratio of the two outgoing currents, are practically independent of the magnetic flux. Therefore, both currents have the same flux dependence (up to a constant multiplicative factor). This explains the ``in-phase" behavior.
 Section \ref{conc} presents our conclusions.

\section{Experiments}\label{exp}

We employed an AB ring connected to a tunnel-coupled wire shown in Fig. 1. This three-terminal geometry is the simplest for realizing the two-slit experiment even compared with the four-terminal geometry employed in the previous work. \cite{naturenano}
Our device is fabricated from a modulation doped AlGaAs/GaAs heterostructure (depth of 2DEG: 125 nm, carrier density: $1.9 \times 10^{11} \ {\rm cm}^{-2}$, mobility: $2 \times 10^{6} \ {\rm cm}^{2}/$Vs) using a standard Schottky gate technique. By varying the tunneling gate voltage $V_{\rm T}$ we can modulate in-situ the tunnel coupling energy $V$. By applying a voltage to the gate $V_{\rm M}$ the phase acquired in one of the two paths can be varied. To observe the quantum interference, a low energy excitation current (excitation energy across the overall sample: 50 $\mu$eV) is injected into the quantum wire on the left, and the output currents $I_\uparrow$ and $I_\downarrow$ are measured simultaneously by sweeping the magnetic field at each gate configuration. All experiments were performed using the dilution refrigerator with a base temperature of 70 mK.

\begin{figure}[ htp]
\hspace{0.5cm} 
\label{figure2}
\includegraphics[width=15cm]{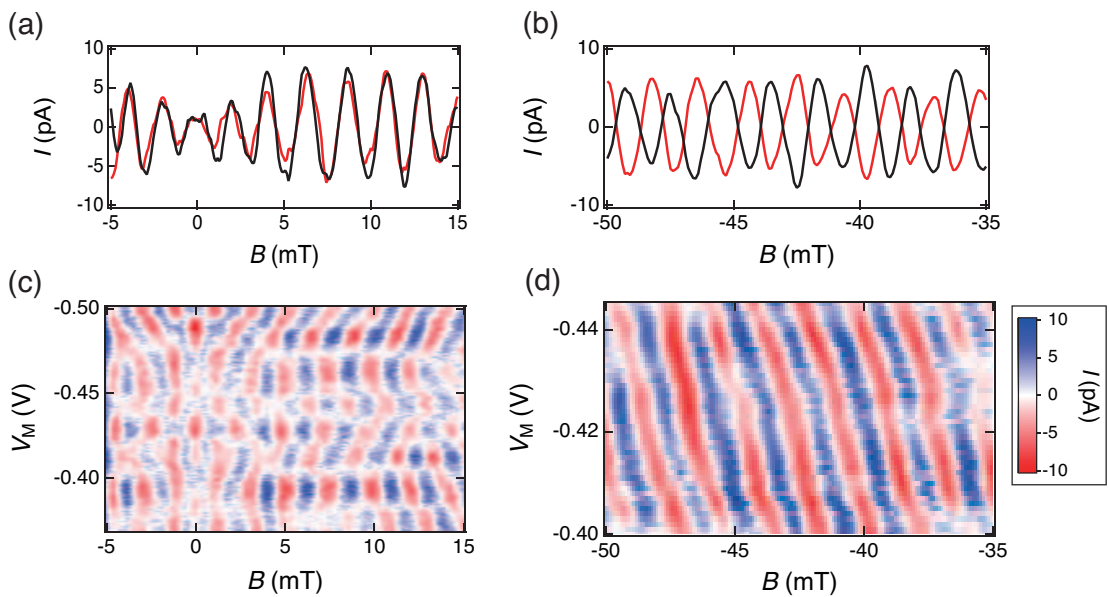}

\caption{(a) Typical AB oscillation in the strong coupling regime. The black and red curves are the measured $I_\uparrow$ and $I_\downarrow$ respectively, after  subtraction of a smoothed background. (b) Typical current oscillation in the weak coupling regime. Tunnel coupling energy is roughly a few 100 $\mu$eV. The oscillations are extracted by subtracting the smoothed background. (c) Intensity plot of $I_\uparrow$ as a function of the perpendicular magnetic field $B$ and side gate voltage $V_{\rm M}$ in the strong coupling regime. (d) Intensity plot of $I_\uparrow$ as a function of $B$ and $V_{\rm M}$ in the weak coupling regime.}
\end{figure}

When we apply a relatively small negative voltage on $V_{\rm T}$, we can deplete the center region of the ring to form an AB ring while keeping strong coupling between the parallel quantum wires. This is because the gate electrode deposited to define the tunnel coupling is narrow. In such a strong coupling case, the two output currents $I_\uparrow$ and $I_\downarrow$ oscillate in-phase as shown in Fig. 2(a). Namely, the two output contacts work equally and the interferometer effectively works as a standard AB ring in a two-terminal setup. The total current oscillates with a period of $h/eS$, where $S$ is the area enclosed by the AB ring. This standard AB interference is subjected to phase rigidity, as would result from the Onsager law. Below we explain why this law applies in this limit.  The phase of the AB oscillation can thus only take the values 0 or $\pi$ at zero magnetic field and as a consequence leads to phase jumps when the AB phase is modulated by changing a voltage $V_{\rm M}$ applied to a side gate of the AB ring as shown in Fig. 2(c). The gate voltage irregularly shifts the phase of the AB oscillation, implying that the observed AB oscillation is not an ideal two-path interference, but a complicated multi-path interference.

In contrast, our device can also be tuned into the weak coupling regime by applying a large negative voltage on $V_{\rm T}$. When $V_{\rm T}$ is properly tuned so that the coupled wire works as a beam splitter to yield high visibility, and the potential change at the transition region between the AB ring and the coupled wire is small enough, the observed two output currents oscillate with opposite phases with almost the same amplitude (see Fig. 2(b)). In other words, the total outgoing current $I_{tot}=I_\uparrow + I_\downarrow$ has very small AB oscillations.  This result strongly suggests that backscattering does not contribute to the main oscillation. Furthermore, when the phase difference between $I_\uparrow$ and $I_\downarrow$ is exactly $\pi$, the phase of the oscillation evolves smoothly and linearly with $V_{\rm M}$ without any jump (see Fig. 2(d)). These results are in contrast to what is observed for the standard two-terminal AB interferometer.
The observed interference does not suffer from the multi-path contribution that modulates the total current, but captures the phase difference between the two paths that linearly shifts with $V_{\rm M}$, suggesting the realization of a  true two-path interference.
 In what follows, we show that the anti-phase oscillation and the smooth phase shift are reproduced in a simple tight-binding model and prove that the measured phase shift is the bare phase shift of the upper path.

 In the experiment, the visibility in both the in-phase and the anti-phase oscillations is further decreased by the existence of many transmitting channels with different tunnel couplings. However, as we show below, a model with a single channel in each wire captures
the above mentioned observed features.

\section{Tight-binding models}\label{models}

We now construct  tight-binding models which imitate the experimental setup and allow a systematic study of the effects of various parameters on the outgoing currents. The models contain single-level sites $n$, with site energies $\epsilon^{}_n$, and nearest-neighbor hopping matrix elements $J^{}_{nm}$. The models are  simple enough to be solved fully analytically, so that we can capture clearly the properties of all eigenstates.  
The Hamiltonian is thus written as
\begin{eqnarray}
&{\cal H}=\sum_{n}\epsilon^{}_n|n\rangle\langle n|-\sum_{\langle nm\rangle}\bigl (J^{}_{nm}|n\rangle\langle m|+{\rm h.c.}\bigr ), \label{eq1}
\end{eqnarray}
where $\langle nm\rangle$ denotes a bond between the neighboring sites $n$ and $m$.
The Schr\"{o}dinger equation for the electron's wave function $|\Psi\rangle\equiv\sum_m \langle m|\Psi\rangle |m\rangle\equiv \sum_m \psi(m)|m\rangle$ is therefore
\begin{eqnarray}
(\epsilon-\epsilon^{}_n)\psi(n)=-\sum_m J^{}_{nm}\psi(m)\ , \label{sch}
\end{eqnarray}
where  $\epsilon$ is the energy of the electron.

\begin{figure}[ hbtp]
\center
(a)  \includegraphics{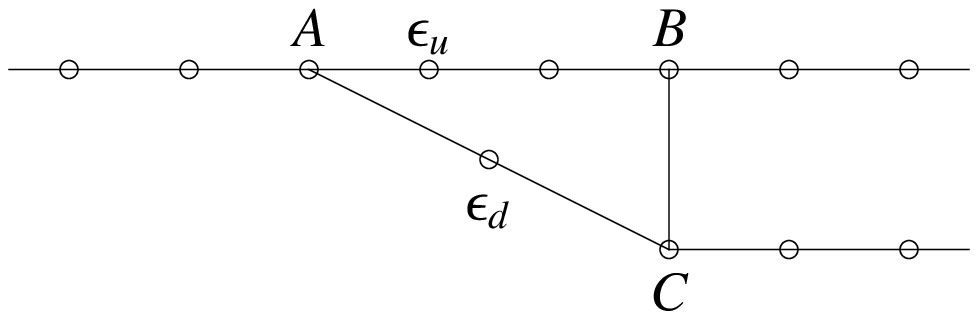}\\
\vspace{1cm}
(b)  \includegraphics{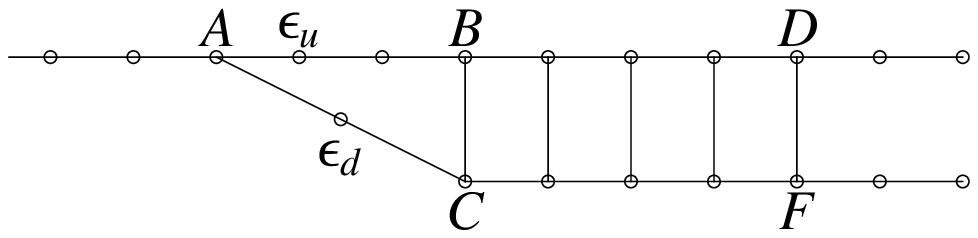}
\caption{(a) The simplest three-terminal model. (b) The model for the interferometer with the tunneling-coupled wires.}
\label{sAB}
\end{figure}

The system is connected to three leads, one on the left hand side and two on the right hand side (see  Fig. \ref{sAB}). These leads are described by one-dimensional chains, with zero site energies and constant nearest-neighbor hopping matrix elements $J^{}_{n,n+1}=J^{}_{n,n-1}\equiv J$. The left lead has $n\leq 0$, and the two outgoing leads have $n\geq n^{}_1$. Within each lead, the Schr\"{o}dinger equation is
\begin{eqnarray}
\epsilon \psi(n)=-J[\psi(n-1)+\psi(n+1)]\ ,
\end{eqnarray}
with the solutions $\psi=e^{\pm ikn}$ and $\epsilon=-2 J \cos(k)$, where the dimensionless wave number $k$ contains the lattice constant. In the following we look for a scattering solution, in which we set
\begin{eqnarray}
&\psi_{in}(n)=e^{ikn}+r e^{-ikn}\ ,\ \ \ \ \ n\leq 0,\nonumber\\
&\psi_{out}^\uparrow(n)=t^\uparrow e^{ik(n-n^{}_1)}\ ,\ \ \ \ \ \ \ \  n\geq n^{}_1\ , \nonumber\\
&\psi_{out}^\downarrow(n)=t^\downarrow e^{ik(n-n^{}_1)}\ ,\ \ \ \ \ \ \ \  n\geq n^{}_1\ . \label{inout}
\end{eqnarray}
Here, $r$ denotes the amplitude of the reflected wave, while $t^\uparrow$ and $t^\downarrow$ denote the amplitudes of the two outgoing waves. These amplitudes determine the two outgoing currents and the reflected current, via the  Landauer formula.\cite{buttiker}
In the linear response limit (zero bias between the left and right leads) and at zero temperature, the ratios of the two outgoing currents and of the reflected current to the incoming one are given by the two transmission and one reflection coefficients,
\begin{eqnarray}
T^\uparrow\equiv|t^\uparrow|^2\ ,\ \ \ T^\downarrow\equiv|t^\downarrow|^2\ ,\ \ \ R=|r|^2\ ,
\end{eqnarray}
with
\begin{eqnarray}
T^\uparrow+T^\downarrow+R=1\ .
\end{eqnarray}

We next start with a simple three-terminal interferometer, Fig. 3(a), for which we give an explicit analytical solution. As we show, this simple case already captures much of the ``anti-phase" behavior. We then add the tunneling-coupled wires, and show how they generate the ``in-phase" behavior for strong tunneling strength.

\section{Three terminal AB interferometer}\label{3term}

The simplest model for a three terminal interferometer is shown in Fig. \ref{sAB}(a). The interferometer is modeled by a triangle $ABC$ of single-level sites. Each corner of this triangle is connected to a one-dimensional lead, as described above.
 On the upper and lower arms of the triangle (AB and AC) we place one-dimensional chains of sites, of lengths $n^{}_u$ and $n^{}_d$, (in the figure, $n^{}_u=3$, $n^{}_d=2$) with uniform site energies  $\epsilon^{}_u$ and $\epsilon^{}_d$ and with uniform nearest-neighbor hopping energies $j^{}_u$ and $j^{}_d$. The vertical arm represents the tunneling between the two channels, with tunneling energy $V$ between the sites $B$ and $C$.

 Within the upper and lower arms, the  solutions to Eqs. (\ref{sch})  are given by
 \begin{eqnarray}
\psi^{}_\ell(n)=\bigl [\sin(k^{}_\ell n)\psi^{}_\ell(n^{}_\ell)+\sin[k^{}_\ell(n^{}_\ell-n)]\psi(A)\bigr ]/\sin(k^{}_\ell n^{}_\ell)\ ,\label{7}
\end{eqnarray}
 where $\ell=u,d$,  and the wave number $k^{}_\ell=\arccos[(\epsilon^{}_\ell-\epsilon)/(2j^{}_\ell)]$ again contains the appropriate lattice constant. In the geometry of Fig. \ref{sAB}(a), we have $\psi^{}_u(0)=\psi^{}_d(0)=\psi(A)=1+r$,
  $\psi^{}_u(n^{}_u)=\psi(B)=t^\uparrow$ and $\psi^{}_d(n^{}_d)=\psi(C)=t^\downarrow$. Gauge invariance allows us to place the AB phase on any bond around the interferometer loop.\cite{AB} Measuring this flux $\Phi$ in units of the unit quantum flux times $2\pi$,
  we place the corresponding phase on the bond BC, and the tight binding equations for the triangle become
 \begin{eqnarray}
&\epsilon t^\uparrow=-Je^{ik}t^\uparrow-Ve^{-i\Phi}t^\downarrow-j^{}_u\psi^{}_u(n^{}_u-1)\ ,\nonumber\\
&\epsilon t^\downarrow=-Je^{ik}t^\downarrow-Ve^{i\Phi}t^\uparrow-j^{}_d \psi^{}_d(n^{}_d-1)\ ,\nonumber\\
&\epsilon(1+r)=-j^{}_u \psi^{}_u(1)-j^{}_d \psi^{}_d(1)-J(e^{-ik}+r e^{ik})\ . \label{eqsn}
\end{eqnarray}
From Eq. (\ref{7}), one has $\psi^{}_u(1)=(1+r)y^{}_u+t^\uparrow x^{}_u$ and $\psi^{}_u(n^{}_u-1)=(1+r)x^{}_u+t^\uparrow y^{}_u$, with similar expressions for the lower branch, with  $x^{}_\ell=j^{}_\ell\sin(k^{}_\ell)/\sin(k^{}_\ell n^{}_\ell),~y^{}_\ell=j^{}_\ell\sin[k^{}_\ell(n^{}_\ell -1)]/\sin(k^{}_\ell n^{}_\ell)$. Finally, one finds the solutions
\begin{eqnarray}
&t^\uparrow=[x^{}_u(Je^{-ik}-y^{}_d)+Vx^{}_d e^{-i\Phi}](1+r)/d\ ,\nonumber\\
&t^\downarrow=[x^{}_d(Je^{-ik}-y^{}_u)+Vx^{}_u e^{i\Phi}](1+r)/d\ ,\nonumber\\
&1+r=-2iJ\sin(k)d/D\ , \label{ttt}
\end{eqnarray}
where
\begin{eqnarray}
&d=(Je^{-ik}-y^{}_u)(Je^{-ik}-y^{}_d)-V^2\ ,\nonumber\\
&D=(Je^{-ik}-y^{}_u-y^{}_d)d-x_u^2(Je^{-ik}-y^{}_d)\nonumber\\
&\ \ \ -x_d^2(Je^{-ik}-y^{}_u)-2x^{}_u x^{}_d V \cos\Phi\ .
\end{eqnarray}
Interestingly, $r$ depends on $\Phi$ only via the term with $\cos\Phi$ in the denominator $D$. Therefore, $R=|r|^2=1-T^\uparrow-T^\downarrow$ is an even function of the flux, as might be expected from the Onsager relation.

To imitate a flat density of states ($d\epsilon/dk$) in the external leads (within the present tight-binding model), one usually chooses electron energies near the center of the band, $\epsilon=0$ or $k=\pi/2$. For this energy, one has
\begin{eqnarray}
&T^\uparrow=4J^2[x_u^2(J^2+y_d^2)+x_d^2V^2+2x^{}_u x^{}_dV(J\sin\Phi-y^{}_d\cos\Phi)]/|D|^2\ ,\nonumber\\
&T^\downarrow=4J^2[x_d^2(J^2+y_u^2)+x_u^2V^2-2x^{}_u x^{}_dV(J\sin\Phi+y^{}_u\cos\Phi)]/|D|^2\ , \label{Tud}
\end{eqnarray}
The denominator has the general form  $D=Q-2x^{}_ux^{}_dV\cos\Phi+i P$, where
\begin{eqnarray}
&P=J(J^2-y_u^2-y_d^2-3y^{}_uy^{}_d+x_u^2+x_d^2+V^2)\equiv P^{}_0+JV^2\ ,\nonumber\\
&Q=(2J^2-y^{}_uy^{}_d+V^2)(y^{}_u+y^{}_d)+x_u^2y^{}_d+x_d^2y^{}_u\equiv Q^{}_0+(y^{}_u+y^{}_d)V^2\ ,
\end{eqnarray}
so that $|D|^2=P^2+(Q-2x^{}_u x^{}_d V\cos\Phi)^2$.

\begin{figure}[ hbtp]
\center
(a)   \includegraphics[width=6cm]{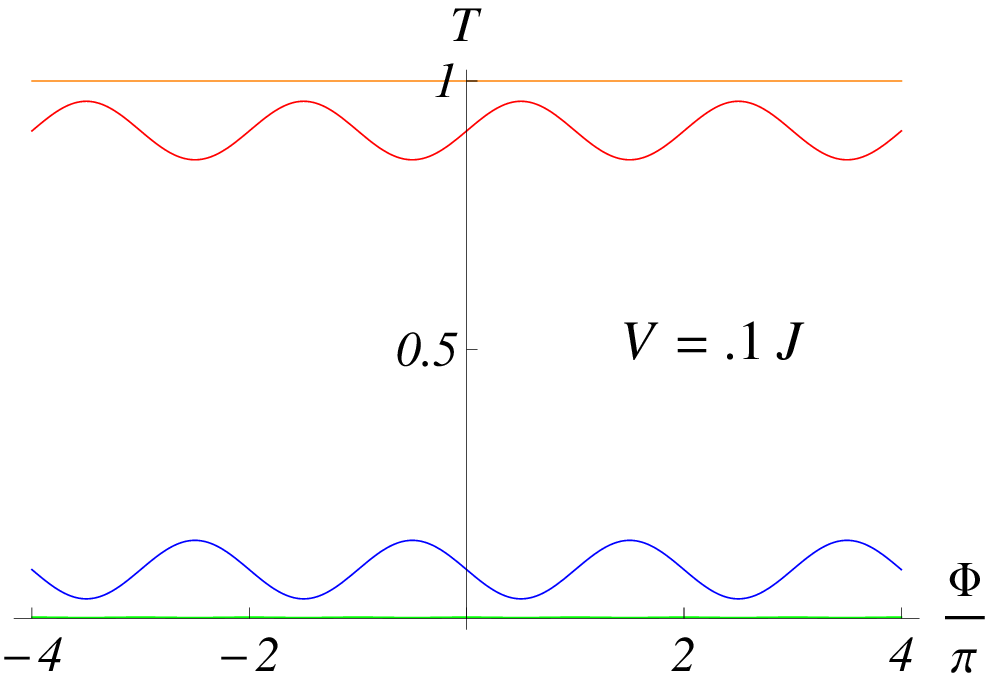}\ \ \ \ \ (b)    \includegraphics[width=6cm]{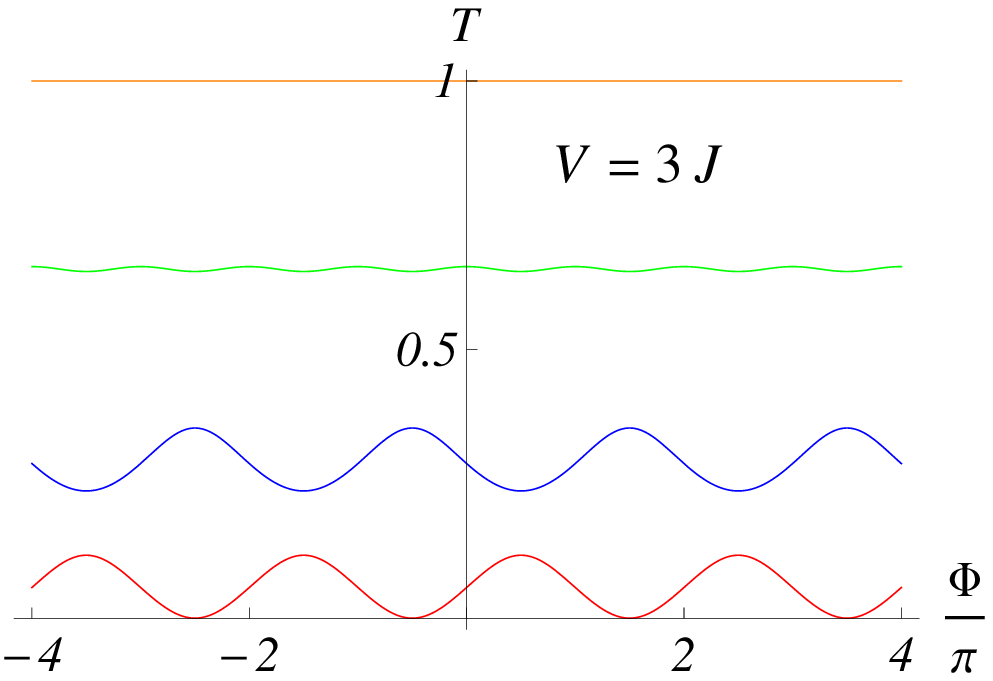}
\caption{Typical results for the simple three-terminal interferometer, with $k=\pi/2$ (i.e. $\epsilon=0$), $j^{}_u=J,~j^{}_d=.3J$, $n^{}_u=n^{}_d=1$. Red: $T^\uparrow$. Blue: $T^\downarrow$. Green: $R$. (a) $V=0.1J$. (b) $V=3J$.}
\label{tri}
\end{figure}

For $n^{}_u=n^{}_d=1$ one has $y^{}_u=y^{}_d=0$, and therefore also $Q=0$, and  $|D|^2=P^2+4x_u^2x_d^2V^2\cos^2\Phi$, while  both numerators in Eqs. (\ref{Tud}) contain the term $\pm 2 x^{}_ux^{}_d J V\sin\Phi$. It turns out that in this special case the numerators determine the locations of the maxima and minima of the two transmissions, and therefore the two outgoing currents  always have opposite phases, with maxima or minima at $\Phi=(1/2+m)\pi$ (with integer $m$) for all the values of the various parameters.  Examples are shown in Fig. \ref{tri}.

\begin{figure}[ hbtp]
\center
(a)   \includegraphics[width=6cm]{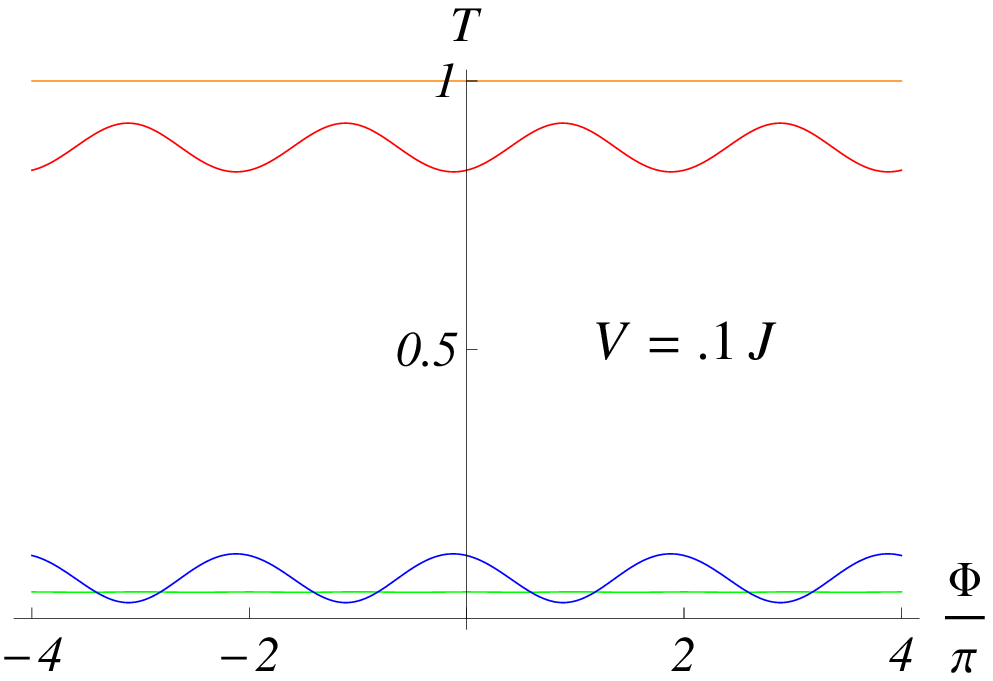}\ \ \ \ \  (b)   \includegraphics[width=6cm]{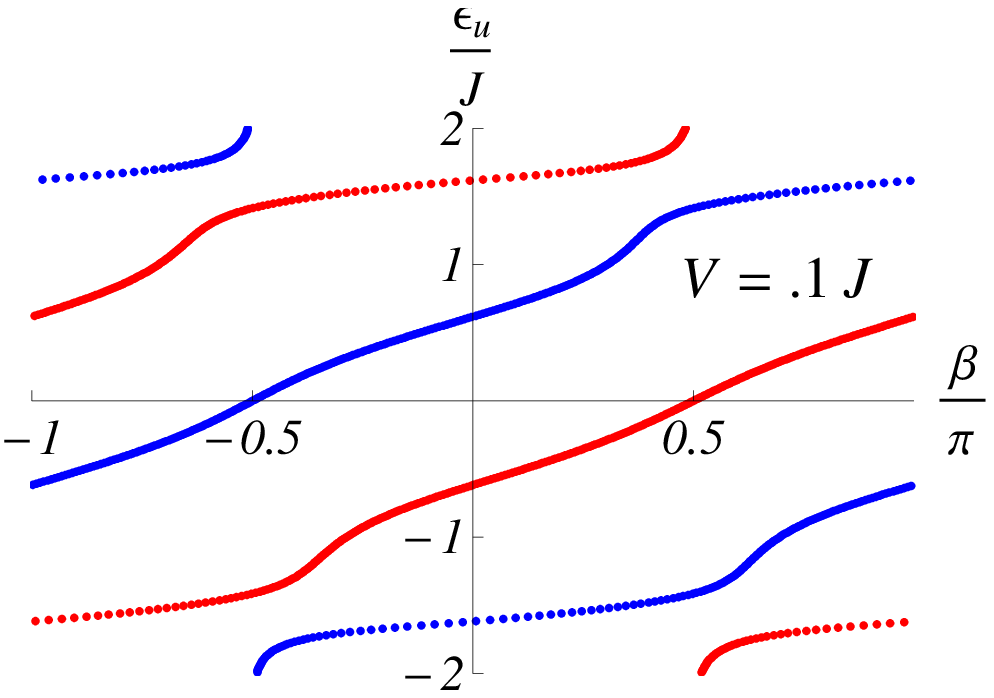}
\caption{(a) Same as Fig. \ref{tri}(a), but with $n^{}_u=5$ and $\epsilon^{}_u=.5J$.   (b) The locations of the maxima (red) and minima (blue) in $T^\uparrow$ (denoted by $\beta$)
 as functions of the ``gate voltage" $\epsilon^{}_u$.}
\label{shift}
\end{figure}

In the experiment it is difficult to tune the interferometer exactly into the symmetric case discussed above. Also,
in the special case $n^{}_u=n^{}_d=1$, the model contains no dependence on the gate voltages on the branches of the AB interferometer; for example, the site energy $\epsilon^{}_u$ is included in the model only for sites between $A$ and $B$ in Fig. \ref{sAB}, which requires $n^{}_u>1$.  To investigate the dependence of the results on the gate voltage $V_M$ (Fig. 1), which is represented by the site energy $\epsilon^{}_u$, we thus studied the model with $n^{}_u>1$. For a small tunnel coupling, $V=0.1J$, typical results are shown in  Fig. \ref{shift}(a). This figure was drawn for $n^{}_u=5$,  a ``gate voltage" on the upper arm of the interferometer $\epsilon^{}_u=.5J$ and $n^{}_d=1$ (so that $x^{}_d=j^{}_d$ and $y^{}_d=0$). As seen in the figure, the oscillations in $R$ have a very small amplitude. Since $T^\uparrow+T^\downarrow=1-R$, this means that the oscillations in the two outgoing transmissions must be in ``anti-phase", as indeed seen in the same figure.

 In the experiments, an anti-phase behavior was also always accompanied by a smooth shift of the maxima and minima of the outgoing currents with the gate voltage $V_M$. To test for this, Fig. \ref{shift}(b) shows the locations of the maxima and minima of $T^\uparrow$ as function of the ``gate voltage" $\epsilon^{}_u$. Indeed, this variation is smooth, and there appear no jumps between maxima and minima.   A similar graph for the extrema of $T^\downarrow$ turns out to be very close to Fig. \ref{shift}(b): the minima of $T^\downarrow$ are very close to the maxima of $T^\uparrow$, and vice versa, for practically all the values of the gate voltage.
  A similar behavior is found for other values of $n^{}_u$. The only difference is that the number of extrema at a fixed flux (in the range  $-2J<\epsilon^{}_u<2J$) is equal to $(n^{}_u-1)$. This number is related to the number of eigenstates within the upper chain (which can be viewed as an extended quantum dot with $(n^{}_u-1)$ levels).
  For small $V$, these results can also be obtained analytically.  Expanding $T^\uparrow$ and $T^\downarrow$ to linear order in $V$ yields
\begin{eqnarray}
&T^\uparrow\approx\frac{4J^2}{P_0^2+Q_0^2}[A^{}_u+2x^{}_ux^{}_dV(J\sin\Phi+C^{}_u\cos\Phi)]\ ,\nonumber\\
&T^\downarrow\approx\frac{4J^2}{P_0^2+Q_0^2}[A^{}_d+2x^{}_ux^{}_dV(-J\sin\Phi+C^{}_d\cos\Phi)]\ , \label{tud}
\end{eqnarray}
where $A^{}_u=x_u^2(J^2+y_d^2)$, $A^{}_d=x_d^2(J^2+y_u^2)$, $C^{}_u=2A^{}_u Q^{}_0/(P_0^2+Q_0^2)-y^{}_d$, $C^{}_d=2A^{}_d Q^{}_0/(P_0^2+Q_0^2)-y^{}_u$, while $P^{}_0$ and $Q^{}_0$ are the values of $P$ and $Q$ at $V=0$.
The last terms in the transmissions (\ref{tud}) can be written as  $J\sin\Phi+C^{}_u\cos\Phi=\sqrt{J^2+C_u^2}\cos(\Phi-\beta^{}_u)$ and $-J\sin\Phi+C^{}_d\cos\Phi=-\sqrt{J^2+C_d^2}\cos(\Phi-\beta^{}_d)$, with
\begin{eqnarray}
\beta^{}_u=\arccos(C^{}_u/\sqrt{J^2+C_u^2}),\ \ \ \ \beta^{}_d=\arccos(-C^{}_d/\sqrt{J^2+C_d^2})\ . \label{beta}
\end{eqnarray}
 Since $C^{}_u$ and $C^{}_d$ have a smooth dependence on the ``gate voltages" $\epsilon^{}_u$ and $\epsilon^{}_d$, this implies a smooth dependence of the observed phase shifts on these energies. In fact, we find numerically that $C^{}_u$ is close to $-C^{}_d$, and therefore $\beta^{}_u\approx \beta^{}_d$, again consistent with the anti-phase behavior.

  One can also expand Eq. (\ref{beta}) to show that when the $\beta$'s are small, they are linear in the the bare phase shifts on the upper arm, e.g. $\beta^{}_u\propto n^{}_uk^{}_u- m\pi$, with $m$ an integer. As a result, they are also linear in $\epsilon^{}_u$.  In this linear regime the phase shift $\beta^{}_u$ is directly proportional to the  shift in the optical path of the electron wave function on the upper branch. A measurement of this phase shift yields this bare phase shift, as in the two-slit interferometer! We have thus demonstrated that our system can be used for measuring phase shifts. Unlike the  open two-terminals interferometer, used for the same purpose by Shuster {\it et al.} \cite{shuster}, our system does not necessarily have a small visibility (i.e.  a small amplitude of the AB oscillations). In practice, the  visibilities in our system are decreased by the existence of many channels with different tunnel coupling.

The above anti-phase behavior, and the smooth variation of the phases of both outgoing currents with the gate voltage, appear only for small $V$. For large $V$, both of the outgoing currents are small, proportional to $1/V^2$, and the reflection $R$ is close to 1. The visibility is even smaller, of order $1/V^3$. In this limit, the details of the AB oscillations depend on the parameters of the device. For example, if the hopping energy through the lower branch of the interferometer is large (e.g. for a large $j^{}_d$, or  near a resonance of the dots on this branch, when $n^{}_d>1$ and $|x^{}_d|,\ |y^{}_d|\gg 1$) we find that $I^\downarrow\ll I^\uparrow\ll R\simeq 1$. In this case, one has $I^\uparrow\simeq 1-R$, and therefore the minima of $I^\uparrow$ follow the maxima of $R$. Since $R$ is even in $\Phi$, these extrema are "phase-locked" at integer multiples of $\pi$, just as for the unitary two-terminal interferometer. If one ignores the small current in the lower outgoing wire then the system indeed behaves like the two-terminal interferometer. Technically, one can see this from Eq. (\ref{Tud}): all the $\Phi-$dependent terms there change sign when $x^{}_u$ changes sign, which happen after crossing resonances in the upper branch.  In contrast, the phases of the small $I^\downarrow$ are not limited by the Onsager restrictions, since most of the current "leaks" though the upper wire. Therefore, from the point of view of the lower wire, the interferometer is `open' \cite{phase,bih}, and the phase of $I^\downarrow$ varies smoothly with the "gate-voltage" $\epsilon^{}_u$. An example of the maxima and minima of both currents in such a case is shown in Fig. 6(a,b). This simple separation between the "two-terminal" (for the upper wire) and the "two-slit" (for the lower wire) behaviors disappears when the two outgoing currents are comparable, see Fig. 6(c,d).

Although the above results imitate many of the experimental results at small $V$, 
the behavior for large $V$ does not reproduce the ``in-phase" behavior observed experimentally  for the tunnel-coupled wires. Therefore, we now turn
 to model the latter system.

\begin{figure}[ hbtp]
\center
(a)  \includegraphics[width=6cm]{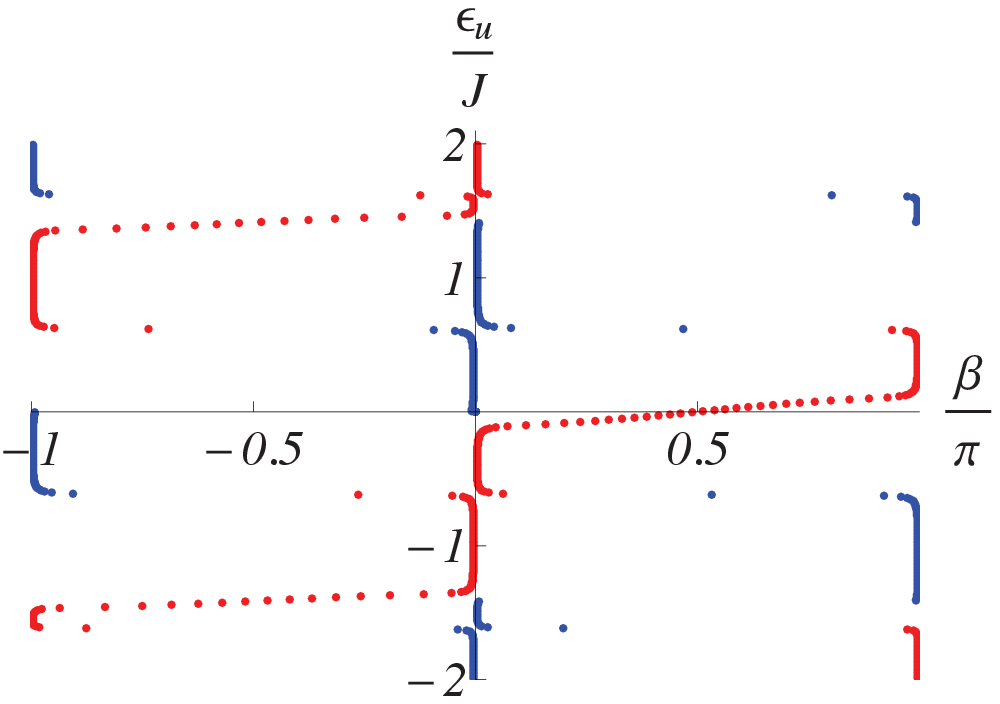}\ \ \ \ (b)\includegraphics[width=6cm]{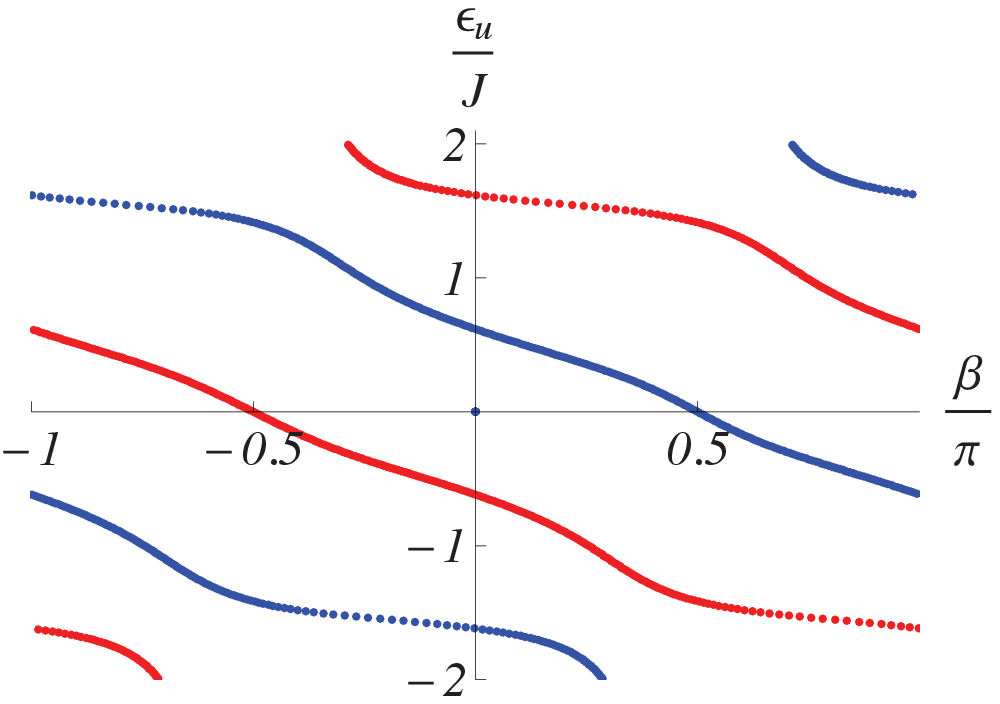}\\
(c)  \includegraphics[width=6cm]{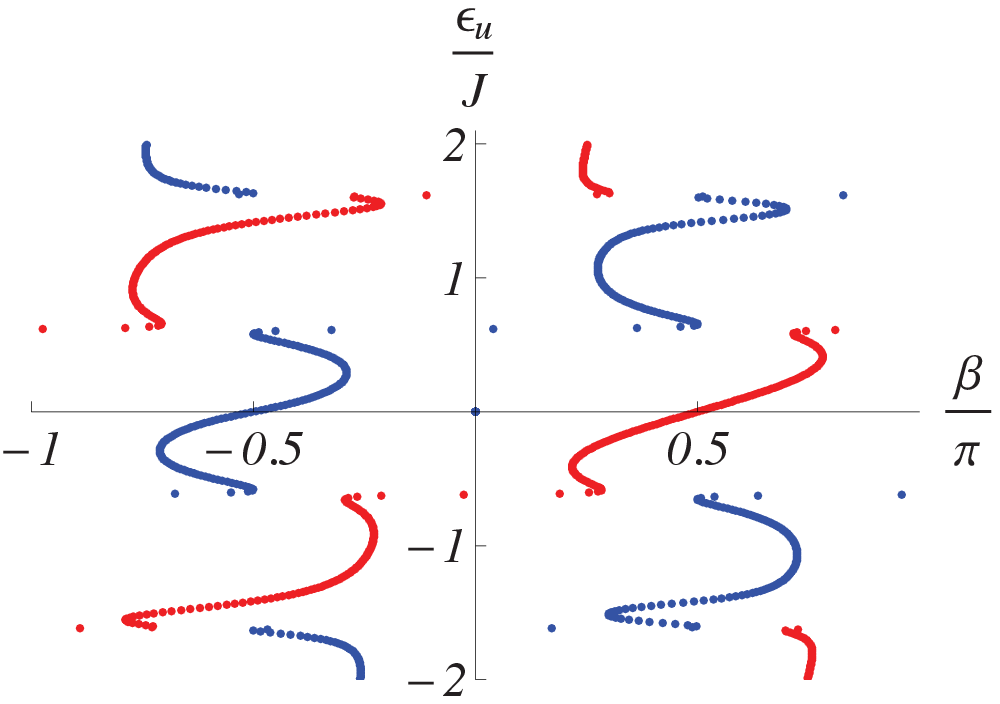}\ \ \ \ (d)\includegraphics[width=6cm]{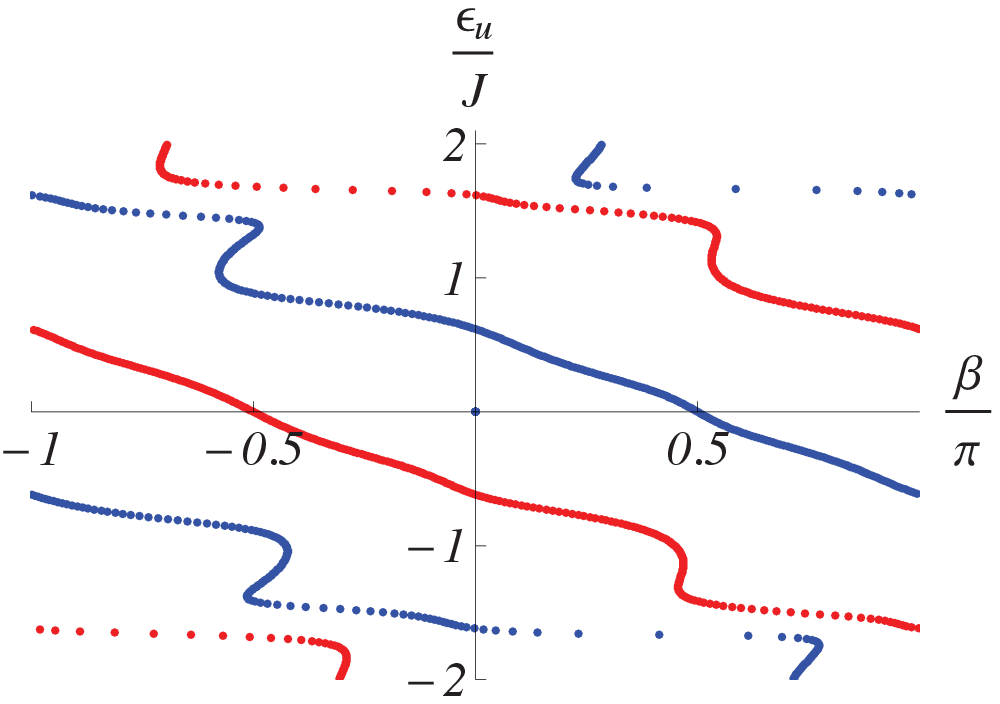}
\caption{ Locations of maxima (red) and minima (blue) for $V=10J$.   The other parameters are the same as in Fig. \ref{shift}.  (a) $T^\uparrow$ for $j^{}_d=100J$. (b) $T^\downarrow$  for $j^{}_d=100J$,
(c) $T^\uparrow$ for $j^{}_d=J$. (d) $T^\downarrow$  for $j^{}_d=J$.}
\label{new3}
\end{figure}

\section{The tunnel-coupled wires}\label{wires}

 Our model for the tunnel-coupled wires is shown in Fig. \ref{sAB}(b).  The triangle ABC on the left still represents the AB interferometer, with a flux $\Phi$ penetrating it. The sequence of $N$ rectangular loops within the rectangle BDFC represents the two coupled wires, BD and CF. Each such loop has a flux $\phi$ (in the same units) through it, and each vertical bond represents the tunneling matrix element $V$ between the wires. Since the confinement of each wire is strong and the transverse spreading of the wave function is much smaller than other length scales of the sample, this simple representation of replacing each quantum wire with a single chain is usually sufficient. In most of the following calculations we imitate the experiment and assume that the area of BDFC is equal to one third of the area of the interferometer loop ABC, and therefore we use $\phi=\Phi/(3N)$. To imitate the continuous tunnel wires one would like $N$ to be very large, and therefore $\phi$ is very small. In practice we perform calculations at several large values of $N$, and ensure that the results do not vary much with $N$.
 The two horizontal lines on the right hand side  represent the two outgoing leads, as before.
The tight binding equations presented in the previous subsection are now supplemented by the equations for the sites on the upper and lower branches of the ladder, which we denote by
 $\psi^\uparrow (n)$ and by $\psi^\downarrow (n)$, respectively ($n=0,~1,~2,...,~N$). Choosing the same gauge as before,  the Schr\"{o}dinger equations for an electron with energy $\epsilon$ on these sites ($0<n<N$), are
\begin{eqnarray}
(\epsilon-\epsilon^{}_\uparrow)\psi^\uparrow (n)&=-j^{}_\uparrow[\psi^\uparrow (n+1)+ \psi^\uparrow (n-1)]-Ve^{-i\phi^{}_n}\psi^\downarrow (n)\ ,\nonumber\\
(\epsilon-\epsilon^{}_\downarrow)\psi^\downarrow (n)&=-j^{}_\downarrow[\psi^\downarrow (n+1)+\psi^\downarrow (n-1)]-Ve^{i\phi^{}_n}\psi^\uparrow (n)\ ,
\end{eqnarray}
where  $\epsilon^{}_\uparrow$ and $\epsilon^{}_\downarrow$ are the (constant) site energies, which  model  gate voltages  applied to each wire separately, while $j^{}_\uparrow$ and $j^{}_\downarrow$ represent the corresponding hopping energies. Also, $\phi^{}_n=\Phi+n\phi$.  We shall later return to the boundary conditions,
\begin{eqnarray}\label{BC}
&\psi^\uparrow(0)=\psi(B)=\psi^{}_u(n^{}_u)\ ,\ \ \ \psi^\downarrow(0)=\psi(C)=\psi^{}_d(n^{}_d)\ ,\nonumber\\
&\psi^\uparrow(N)=\psi(D)=t^\uparrow\ ,\ \ \ \psi^\downarrow(N)=\psi(F)=t^\downarrow\ .
\end{eqnarray}

\begin{figure}[ hbtp]
\center
\includegraphics[width=8cm]{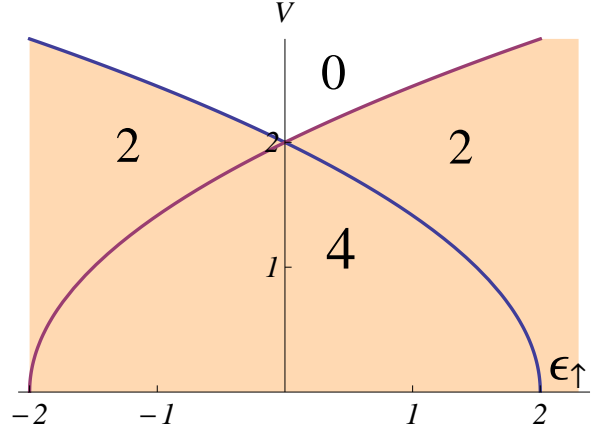}\\
\caption{Regions in the $\epsilon^{}_\uparrow-V$ plane (both in units of $J$) with 4, 2 and no  running solutions; the numbers of running solutions are indicated. Other parameters are $\phi=0$, $j^{}_\uparrow=j^{}_\downarrow=J$, $\epsilon=\epsilon^{}_\downarrow=0$.}
\label{diag}
\end{figure}

We first discuss the solution of the above tight-binding equations within the ladder.
A wave-like solution of these equations can be found by setting $\psi^{\uparrow}(n)=e^{i (K n-\phi^{}_n/2)}u^{\uparrow}$, $\psi^{\downarrow}(n)=e^{i (K n+\phi^{}_n/2)}u^{\downarrow}$ (where again the dimensionless wave number $K$ contains the lattice constant along the ladder). The amplitudes $u^{\uparrow,\downarrow}$ must then obey the linear equations
\begin{eqnarray}
&[\epsilon-\epsilon^{}_\uparrow+2 j^{}_\uparrow\cos(K-\phi/2)]u^\uparrow+Vu^\downarrow=0\ ,\nonumber\\
&Vu^\uparrow+[\epsilon-\epsilon^{}_\downarrow+2 j^{}_\downarrow\cos(K+\phi/2)]u^\downarrow=0\ .
\end{eqnarray}
Therefore, the wave-numbers $K$ are the solutions of the determinant equation,
\begin{eqnarray}
[\epsilon-\epsilon^{}_\uparrow+2 j^{}_\uparrow\cos(K+-\phi/2)][\epsilon-\epsilon^{}_\downarrow+2 j^{}_\downarrow\cos(K+\phi/2)]-V^2=0\ ,\label{quad}
\end{eqnarray}
i.e.
\begin{eqnarray}
&4j^{}_\uparrow j^{}_\downarrow\cos^2K+2[(\epsilon-\epsilon^{}_\uparrow)j^{}_\downarrow+(\epsilon-
\epsilon^{}_\downarrow)j^{}_\uparrow]\cos(\phi/2)\cos K
\nonumber\\&+(\epsilon-\epsilon^{}_\uparrow)(\epsilon-\epsilon^{}_\downarrow)-V^2-4j^{}_\uparrow j^{}_\downarrow\sin^2(\phi/2)\nonumber\\
&=2[(\epsilon-\epsilon^{}_\uparrow)j^{}_\downarrow-(\epsilon-\epsilon^{}_\downarrow)j^{}_\uparrow]\sin(\phi/2)\sin K\ .\label{eqq}
\end{eqnarray}
As stated, we need results for large $N$ and therefore for small $\phi$.  At $\phi=0$ we have a quadratic equation,
with two solutions, $\cos K=c^{}_\pm\equiv [(\epsilon^{}_\uparrow-\epsilon)/j^{}_\uparrow+(\epsilon^{}_\downarrow-\epsilon)/j^{}_\downarrow\pm\sqrt{[(\epsilon^{}_\uparrow-
\epsilon)/j^{}_\uparrow-(\epsilon^{}_\downarrow-\epsilon)/j^{}_\downarrow]^2+4V^2/(j^{}_\uparrow j^{}_\downarrow)}]/4$.   For small $V$, the two solutions for $\cos K$ remain in the range $-1<c^{}_\pm <1$, and therefore each of them corresponds to waves running in opposite directions, with wave numbers $K^{}_{1,2}=\pm\arccos[c^{}_+]$ and $K^{}_{3,4}=\pm\arccos[c^{}_-]$.  However, as $V$ increases one of $|c^{}_\pm|$ crosses the value $1$  when
$V^2=(\epsilon-\epsilon^{}_\uparrow\pm 2 j^{}_\uparrow)(\epsilon-\epsilon^{}_\downarrow\pm 2 j^{}_\downarrow)$. Above these values of $V$, $K^{}_{1,2}$ and/or $K^{}_{3,4}$ become complex, and the corresponding waves become evanescent.
Figure \ref{diag} shows an example of regions in the $\epsilon^{}_\uparrow-V$ plane, for the special parameters $j^{}_\uparrow=j^{}_\downarrow=J$, and $\epsilon=\epsilon^{}_\downarrow=0$. The numbers on the diagram (4, 2 or 0) indicate the number of ``running" solutions, with real $K$'s, in each region.

For a large but finite $N$ we need to solve Eq. (\ref{eqq}) for a finite small $\phi$. In practice we do that by searching a solution for $K$ close to each of the four solutions found at $\phi=0$. For each value of $\phi$ this yields four waves, with wave numbers $K^{}_1,~K^{}_2,~K^{}_3,~K^{}_4$ [an alternative, which gives the same results, is to square Eq. (\ref{eqq}), solve a quartic equation for the four solutions $\cos K^{}_\ell$ and then choose the right signs of $\sin K^{}_\ell$ which satisfiy Eq. (\ref{eqq})]. As $V$ increases at fixed $|\epsilon^{}_\uparrow-\epsilon|<2j^{}_\uparrow$, at first all the four $K$'s are real, then  two of them become complex  and then the other two also become complex. As we shall see below, each of these regimes ends up with a different qualitative behavior of the two outgoing currents.

\begin{figure}[ hbtp]
\center
\includegraphics[width=10cm]{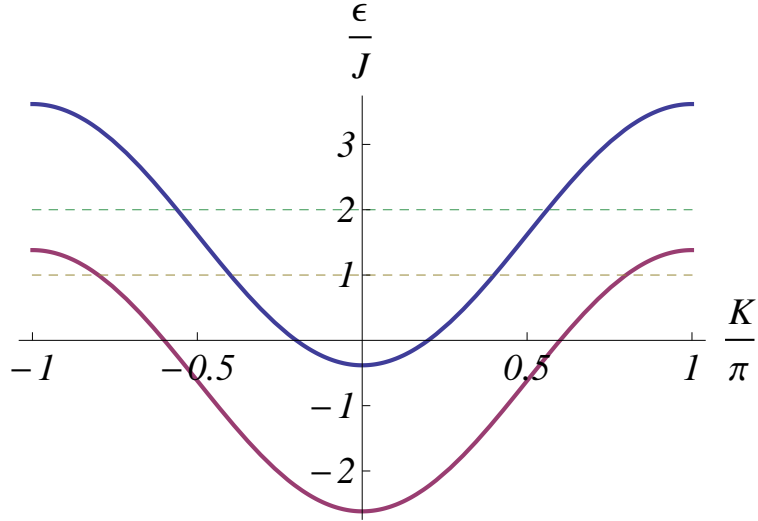}\\
\caption{The two energy bands for the infinite ladder, for $\phi=0$, $j^{}_\uparrow=j^{}_\downarrow=J$, $\epsilon^{}_\downarrow=0$ and  $\epsilon^{}_\uparrow=V=J$.
The dashed lines are at energies with four running solutions ($\epsilon=J$) and two running solutions ($\epsilon=2J$).}
\label{en}
\end{figure}

 For each of the four $K$'s, the corresponding amplitudes of the wave functions obey the relation
\begin{eqnarray}
u^\downarrow_\ell=-u^\uparrow_\ell[\epsilon-\epsilon^{}_\uparrow+2 j^{}_\uparrow\cos(K^{}_\ell -\phi/2)]/V\ .\label{uud}
\end{eqnarray}
The above solutions represent the eigenstates of the infinite periodic ladder. For the infinite ladder, one cannot accept the evanescent solutions, which increase to infinity for $n \rightarrow \infty$ or for $n \rightarrow -\infty$. Therefore, one considers only real values of $K$, and one ends up with two energy bands with energies $\epsilon(K)$ which are given by the solution of the quadratic equation (\ref{eqq}) in $\epsilon$. For the simple case presented in Fig. \ref{diag}, $j^{}_\uparrow=j^{}_\downarrow=J$, $\epsilon=\epsilon^{}_\downarrow=0$ and $\phi=0$, these two bands are given by $\epsilon=[\epsilon^{}_\uparrow-4J\cos K\pm\sqrt{\epsilon_\uparrow^2+4V^2}]/2$, see Fig. \ref{en}. The ``running" solutions at energy $\epsilon$ are found as the intersections of the horizontal line at that energy  with these two functions (see dashed lines in the figure).
At $\epsilon^{}_\uparrow=V=0$ the two bands coincide, and therefore every $\epsilon$ corresponds to two degenerate running waves, namely four waves. However, non-zero values of $\epsilon^{}_\uparrow$ and/or of $V$ split the two bands, and then there exist energy ranges in which there appear only two running solutions, or even gaps for which there are no running solutions. In these regions the remaining waves are evanescent, which are forbidden for the infinite ladder but  allowed in the finite ladder.
As we show below, one needs to be in these regions (and therefore to have some asymmetry between the wires) in order to obtain the in-phase behavior of the outgoing currents.

\section{Interferometer with tunnel-coupled wires}\label{Intwires}

We now return to the model of Fig. \ref{sAB}(b). Using Eq. (\ref{uud}), the general solution for the $N+1$ sites ($n=0,~1,~2,\cdots,~N$) on the $N$ squares of the ladder is  written as a linear combination of the four solutions,
\begin{eqnarray}
\psi^\uparrow(n)&=\sum_{\ell=1}^4 A^{}_\ell e^{i(K^{}_\ell n-\phi^{}_n/2)}\ ,\nonumber\\
\psi^\downarrow(n)&=-\sum_{\ell=1}^4 [\epsilon-\epsilon^{}_\uparrow+2 j^{}_\uparrow\cos(K^{}_\ell-\phi/2)] A^{}_\ell e^{i(K^{}_\ell n+\phi^{}_n/2)}/V\ , \label{eqq3}
\end{eqnarray}
with the four yet unknown amplitudes $\{A^{}_\ell\}$.

Using the analogs of Eqs. (\ref{inout}) on the leads, we now have $t^{\uparrow,\downarrow}\equiv\psi^{\uparrow,\downarrow}(N)$ [see Eqs. (\ref{BC})].
 The Schr\"{o}dinger equations at these two end points (near the outgoing leads) are thus
\begin{eqnarray}
&(\epsilon-\epsilon^{}_\uparrow)\psi^{\uparrow}(N)=-
j^{}_{\uparrow}\psi^{\uparrow}(N-1)-Jt^{\uparrow} e^{ik}-Ve^{-i\phi^{}_N}\psi^{\downarrow}(N)\ ,\nonumber\\
&(\epsilon-\epsilon^{}_{\downarrow})\psi^{\downarrow}(N)=-
j^{}_{\downarrow}\psi^{\downarrow}(N-1)-Jt^{\downarrow} e^{ik}-Ve^{i\phi^{}_N}\psi^{\uparrow}(N)\ . \label{eqq1}
\end{eqnarray}
Similarly, Eqs. (\ref{eqsn}) are now replaced by
the equations for the wave functions at the three corners of the AB interferometer,
\begin{eqnarray}
&\epsilon(1+r)=-J(e^{-ik}+re^{ik})-j^{}_u \psi^{}_u(1)-j^{}_d\psi^{}_d(1)\ ,\nonumber\\
&(\epsilon-\epsilon^{}_\uparrow)\psi^\uparrow(0)=-j^{}_u\psi^{}_u(n^{}_u-1)-j^{}_\uparrow \psi^\uparrow(1)-Ve^{-i\Phi}\psi^\downarrow(0)\ ,\nonumber\\
&(\epsilon-\epsilon^{}_\downarrow)\psi^\downarrow(0)=-j^{}_d\psi^{}_d(n^{}_d-1)-j^{}_\downarrow \psi^\downarrow(1)-Ve^{i\Phi}\psi^\uparrow(0)\ . \label{eqq2}
\end{eqnarray}
Equations (\ref{eqq1}) and (\ref{eqq2}), together with Eqs. (\ref{eqq3}) and (\ref{BC}), now reduce to five linear equations in the five unknowns $\{A^{}_\ell\}$ and $r$. Their solution yields the reflection coefficient $R=|r|^2$ and the two transmission coefficients $T^\uparrow=|t^\uparrow|^2$ and $T^\downarrow=|t^\downarrow|^2$.

As stated, we set $\phi= \Phi/(3N)$. Below we plot results as a function of the flux through the AB loop, namely $\Phi$. For reasonable values of $\Phi$ this implies relatively small values of $\phi$. For the results presented below we used $N=1001$ and $N=401$ (the numerical solutions of the five linear equations become difficult for large $N$ and large $V$, when the factors $e^{i K N}$ vary by many orders of magnitude) . These results change only slightly (quantitatively but not qualitatively) when we used other (large) values of $N$.

In order to observe the dependence of the results on the gate voltage on the upper arm of the interferometer, we again need to have $n^{}_u>1$. Below we present typical results with $n^{}_u=5$. The results do not change qualitatively for a wide range of the other parameters. For small $V$ we always find ``anti-phase" behavior, even when the two wires are symmetric. As explained below,  to see the "in-phase" behavior we need to have only two `running' waves, and this happens only with some asymmetry between the wires (see Fig. \ref{diag}). Indeed, with some such anisotropy and and  for appropriately chosen values of large  $V$ (explained below) we find ``in-phase" behavior. This behavior requires the coupled wires, and did not appear in the simpler three-terminal interferometer presented in Sec. IV.

\begin{figure}[ hbtp]
\center
(a)  \includegraphics[width=6cm]{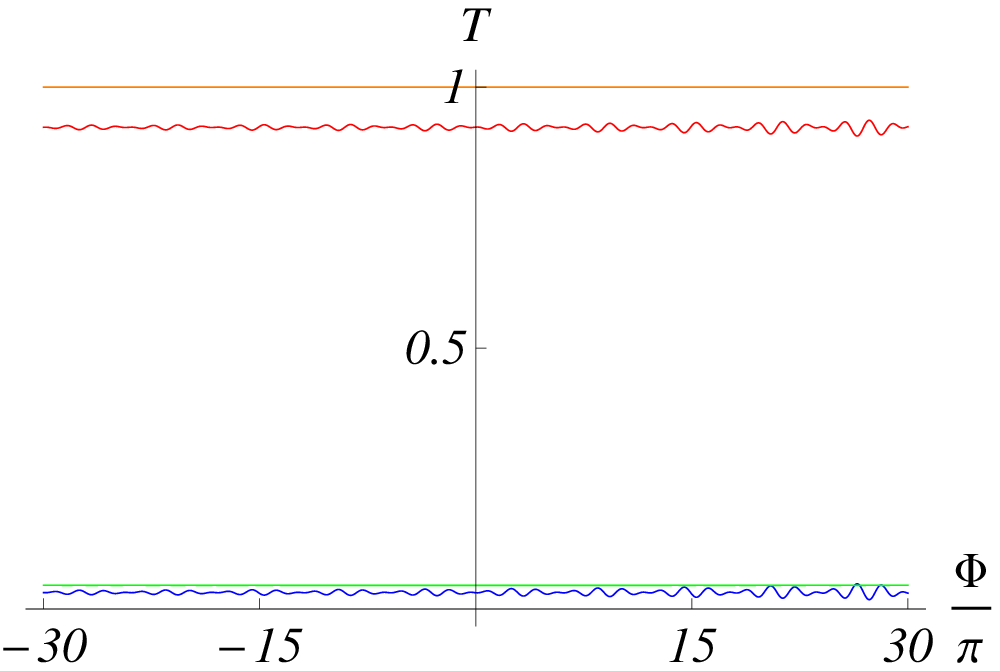}\ \ \ \ (b)  \includegraphics[width=6cm]{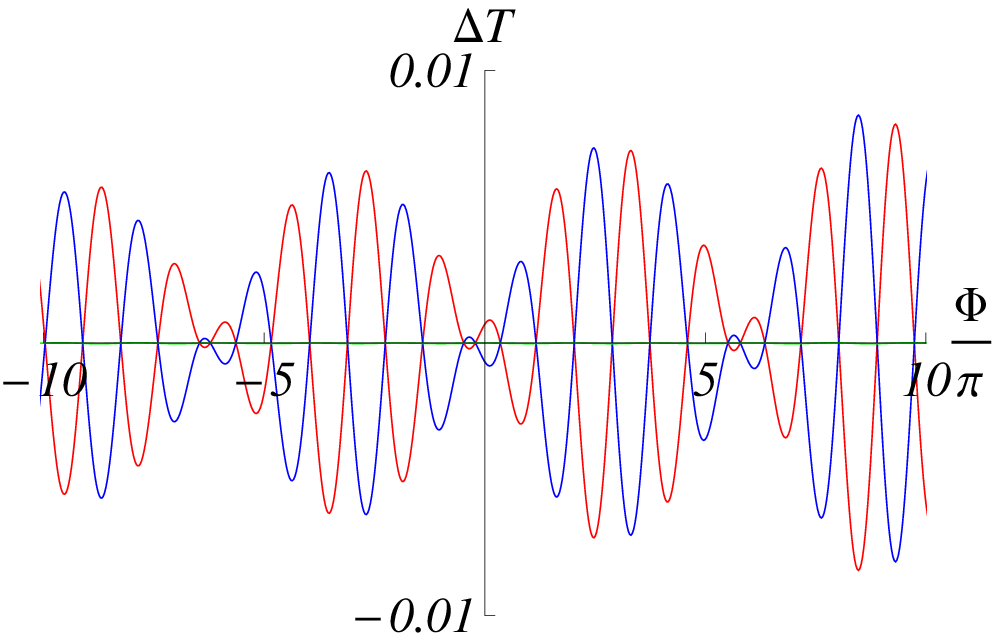}\\

(c)  \includegraphics[width=6cm]{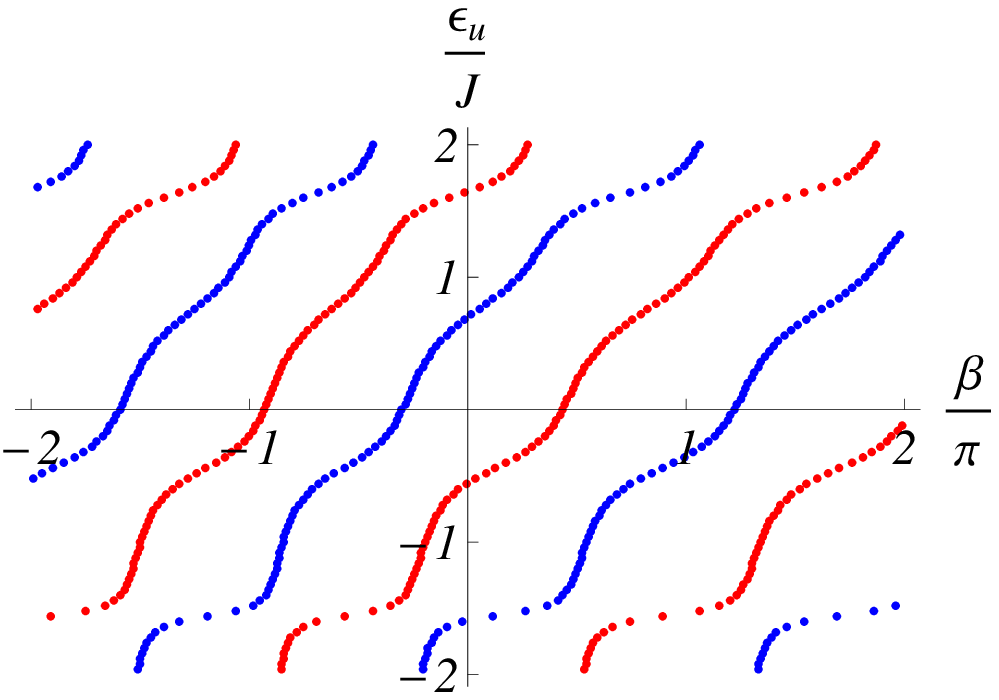}\\
\caption{Typical results for the tunneling-coupled wires, for weak coupling:  $N=1001$, $\phi=\Phi/(3N)$, $\epsilon^{}_\uparrow=.1J$, $\epsilon^{}_\downarrow=0$,  $j^{}_u=j^{}_\uparrow=j^{}_\downarrow=J,~j^{}_d=.2J$, $n^{}_u=5,~n^{}_d=1,~\epsilon=0$ and $V=0.001J$.
(a)  Transmissions $T^\uparrow$ (red) and $T^\downarrow$ (blue) and reflection $R$ (green), for $~\epsilon^{}_u=.1J$.
 (b) An enlarged version of (a), showing $T^\uparrow-\langle T^\uparrow\rangle$, $T^\downarrow-\langle T^\downarrow\rangle$ and
 $R-\langle R\rangle$.  
 (c) Maxima (red) and minima (blue) of $T^\uparrow$ versus $\epsilon^{}_u$ [all other parameters are the same as in (a)].}
\label{new}
\end{figure}

We start with weak coupling, $V=0.001J$. Figure \ref{new}(a) shows results for a small coupling $V$, for the parameters as indicated. Due to the anisotropy between the two branches of the interferometer, $j^{}_u=J,~j^{}_d=.2J$, $T^\uparrow$ is much larger than $T^\downarrow$. However, when we shift each of the transmissions by their average value, as shown in Fig. \ref{new}(b), it is obvious that the AB oscillations of the two transmissions exhibit the same kind of ``anti-phase" behavior as we already saw in the previous section (and as seen experimentally). The main reason for the ``anti-phase"
behavior can be attributed to the weak oscillations in $R$ [which are practically zero in Fig. \ref{new}(b)]. The relation $T^\uparrow+T^\downarrow=1-R$ then requires that the oscillating terms in the two transmissions must have opposite signs, so they cancel each other in the sum. 

Figure \ref{new}(c) shows the variation of the minima and maxima of $T^\uparrow$ with $\epsilon^{}_\uparrow$. Similar to Fig. \ref{shift}(b), these maxima and minima move smoothly with the gate voltage, without any jumps. These maxima and minima again reflect the energy levels of the upper arm AB of the interferometer. Therefore, we conclude that here also the oscillations in the two transmissions reflect the basic ``two-slit" interference around the AB loop.

\begin{figure}[ hbtp]
\center
(a)  \includegraphics[width=6cm]{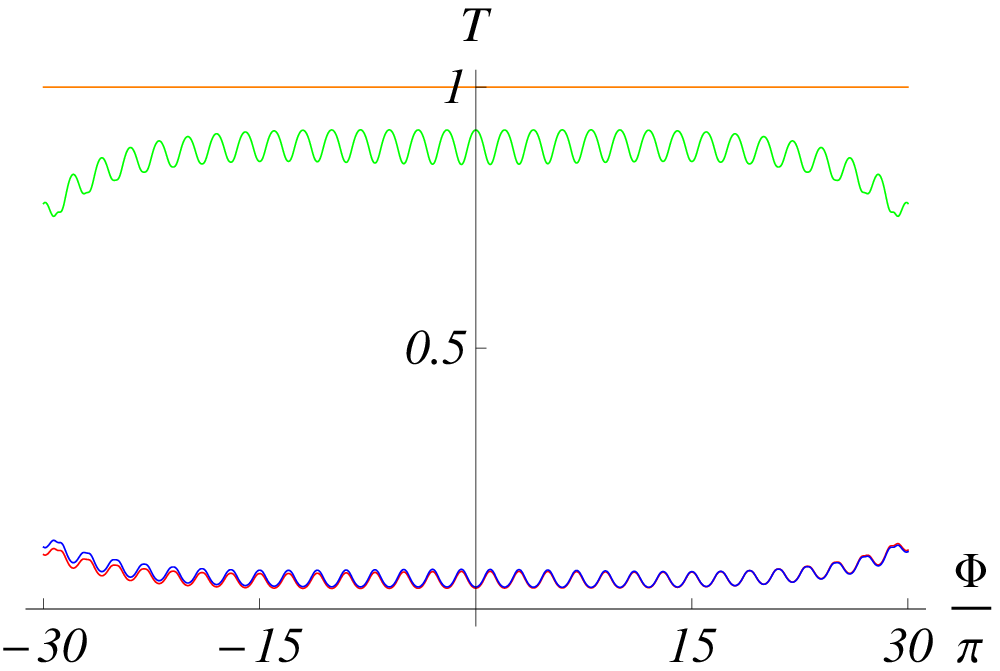}\ \ \ \ (b)  \includegraphics[width=6cm]{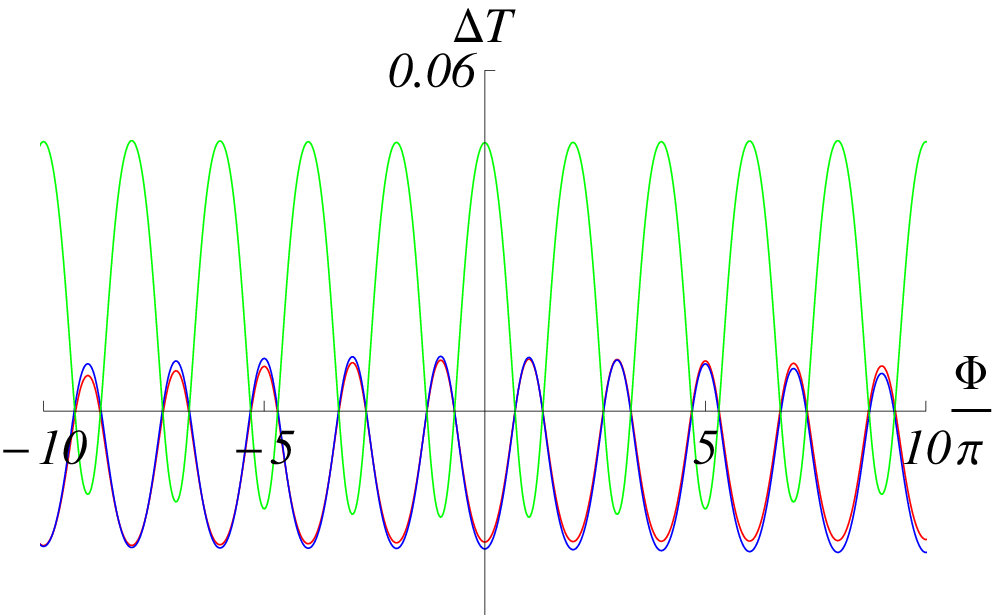}\\
(c)  \includegraphics[width=6cm]{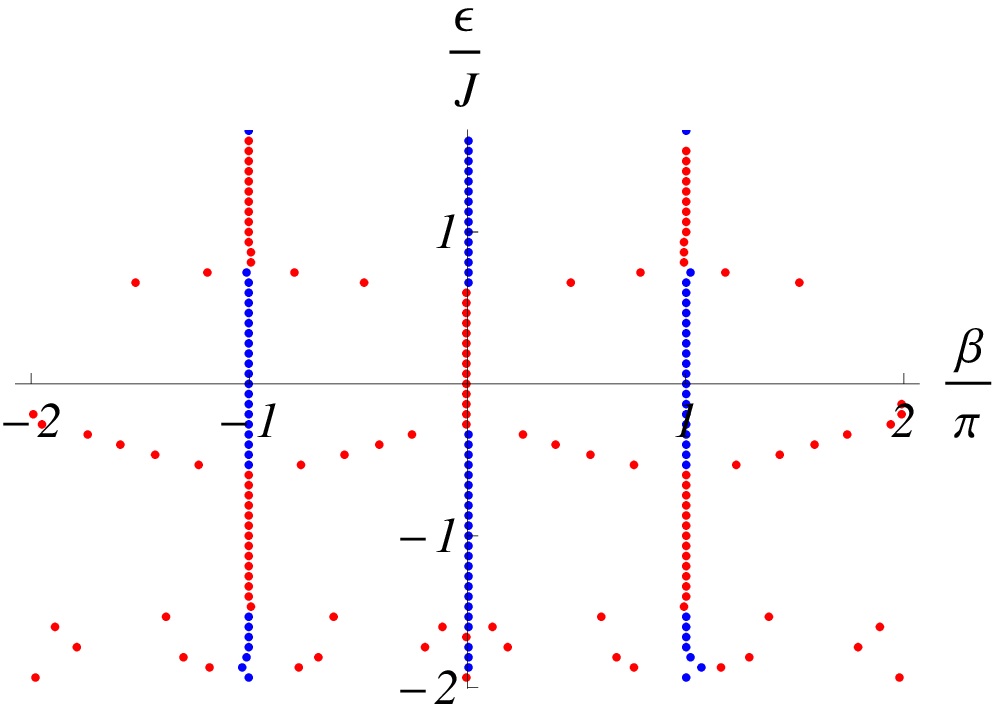}\\
\caption{Typical results for the tunneling-coupled wires, for strong coupling:  $N=401$, $V=1.95J$. All the other parameters are the same as in Fig. \ref{new}.
(a)  Transmissions $T^\uparrow$ (red) and $T^\downarrow$ (blue) and reflection $R$ (green), for $~\epsilon^{}_u=.1J$.
 (b) An enlarged version of (a), showing $T^\uparrow-\langle T^\uparrow\rangle$, $T^\downarrow-\langle T^\downarrow\rangle$ and $R-\langle R\rangle$.  
 (c) Maxima (red) and minima (blue) of $T^\uparrow$ versus $\epsilon^{}_u$ (all other parameters are the same as in (a)).}
\label{new195}
\end{figure}

We now turn to large $V$.  Figure \ref{new195} is similar to Fig. \ref{new}, except that $V=1.95J$ and $N=401$. 
As can be seen from Figs. \ref{new195}(a) and (b), the two outgoing transmissions are practically identical to each other, and therefore they are fully in phase with each other. Furthermore, they both are symmetric under $\Phi \leftrightarrow -\Phi$. As required by the relation $T^\uparrow+T^\downarrow=1-R$, the reflection coefficient $R$ exhibits opposite oscillations, with a double amplitude [see Fig. \ref{new195}(b)].  
The maxima and minima of $T^\uparrow$, shown in Fig. \ref{new195}(c), exhibit full ``phase locking"; they remain at integer multiples of $\pi$. (For some values of $\epsilon^{}_\uparrow$ one observes additional red points between these integer values; these indicate the  splitting of the maxima into pairs of maxima, due to higher integer Fourier components. One always finds only integer Fourier components, but with amplitudes which vary with the parameters).  This phase locking  is similar to that observed in the two-terminal interferometer. \cite{phase} Apart from being locked, maxima occasionally turn into minima and vice versa. This also appeared in Fig. \ref{new3}(a), and must arise from the same origin: crossing of resonances on the branch AB of the interferometer.  All of these features are very similar to those seen experimentally.

For large $V$, electrons are strongly reflected from the points B and C of the structure. This explains the large value of $R$, and the small values of the transmitted currents. These features also appeared for the three terminal interferometer, without the tunnel coupling between the outgoing wires. 
To understand the other features observed in Fig. \ref{new195}, we note that the point $\epsilon^{}_\uparrow=0.1J,~V=1.95V$ is just above the lower line in Fig. \ref{diag} (it is difficult to solve the linear equations for larger $V$, when the imaginary part of the evanescent wave numbers are large and the corresponding wave functions go beyond the accuracy limits of the computer. This is why we stay close to this line, and also why we only present numbers for $N=401$).  Within the region between the two lines in Fig. \ref{diag}, two of the four solutions within the ``ladder" decay exponentially, and we are left with only two running waves, which result from one of the solutions $c^{}_+$ or $c^{}_-$. For $\epsilon^{}_\uparrow>0$, the first evanescent waves (as we cross the lower line in Fig. \ref{diag}) are associated with $c^{}_+$ becoming larger than 1. For the parameters in that figure, this happens at $V^2=2J(2J-\epsilon^{}_\uparrow)$, namely at $V\approx 1.949J$. At that point one has $c^{}_-=\epsilon^{}_\uparrow/(2J)-1\approx -.95$,  and the two remaining running waves have very close values of $|K|$. At the same neighborhood, Eq. (\ref{uud})
yields
\begin{eqnarray}
u^\downarrow/u^\uparrow&=-[\epsilon-\epsilon^{}_\uparrow+2 j^{}_\uparrow\cos(K -\phi/2)]/V\nonumber\\
&\approx -[\epsilon-\epsilon^{}_\uparrow+2 j^{}_\uparrow c^{}_-)]/V\ ,
\label{uud1}
\end{eqnarray}
where we neglected the small $\phi$. Since this ratio applies to both the running solutions, and since the evanescent solutions can be neglected, we conclude that the same ratio applies to their linear combinations, $\psi^\uparrow(n)$ and $\psi^\downarrow(n)$. In particular,
\begin{eqnarray}
T^\downarrow/T^\uparrow= |t^\downarrow/t^\uparrow|^2\approx |\epsilon-\epsilon^{}_\uparrow+2 j^{}_\uparrow c^{}_-|^2/V^2\ ,
\end{eqnarray}
{\it practically independent of $\Phi$} (at least for fluxes which are not too large). Since this ratio is independent of $\Phi$, the two transmissions are proportional to each other, and therefore they are {\it exactly in phase}.  Deviations will occur only at large fluxes, when $\phi$ becomes significant.

For the parameters used in Fig. \ref{new195}, this equation yields $T^\downarrow/T^\uparrow\approx 1/[1-\epsilon^{}_\uparrow/(2J)]\approx 1.05$. This explains the practical overlap of the two transmissions. For other parameters the ratio need not be so close to unity, but the in-phase behavior will persist whenever we have two evanescent waves.
The Onsager relation requires that the full current through the system should be an even function of the flux. This implies that both $R$ and $T^\uparrow+T^\downarrow$ should be such even functions. Indeed, all our calculations show that $R$ is even in $\Phi$. Once we demonstrated that the two transmissions are proportional to each other, each of them must be proportional to their sum, and therefore each of them separately must be an even function of $\Phi$. Combined with the periodicity in $\Phi$, this explains the phase rigidity of the results.

\section{Conclusion}\label{conc}

We have demonstrated a novel two-slit experiment, using an AB ring connected to a tunnel-coupled wire in a three-terminal setup.
Out simple tight binding model captures most of the observed behavior of the currents in the three terminal interferometer. For small tunnel-coupling $V$, one always has the anti-phase behavior. In that limit, the dependence of the the phase of the Aharonov-Bohm oscillations in the outgoing currents on the gate voltage acting on one arm of the interferometer follows  the phase of the electronic wave function on that arm. This three terminal interferometer thus behaves like the two-slit or open interferometer.  For large $V$ one needs to tune the inter-wire coupling to a regime where one has only two running wave solutions. In that regime, the ratio between the two outgoing currents is practically flux independent, and therefore they are in phase, with phase rigidity.
Although our model does not include all the details of the interferometer, such as finite widths of the quantum wires, influence of the multiple transport channels and accompanying electron-electron interaction, the simple and analytically solvable model provides the guiding principles for  realizing a `two-slit' experiment and for a reliable phase measurement in the three-terminal setup.

\ack

Work at Ben Gurion University was supported by the Israel Science Foundation.
S. Takada acknowledges support from JSPS Research Fellowships for Young Scientists. M.Y. acknowledges financial support by Grant-in-Aid for Young Scientists A (no. 23684019). S.Tarucha acknowledges financial support by MEXT KAKENHHI “Quantum Cybernetics", MEXT project for Developing Innovation Systems, and JST Strategic International Cooperative Program. The high quality 2DEG wafer was provided by A. D. Wieck.

\end{document}